\documentclass[notitlepage,aps,nofootinbib,letterpaper,prx,twocolumn,showpacs,amsmath,amssymb,amsfonts,long,superscriptaddress,preprintnumbers,longbibliography]{revtex4-1}

\usepackage{amsmath,amssymb,amsthm,amsxtra,overpic,bbm,bm,epsfig,subfigure}
\usepackage{simplewick} 
\usepackage[dvipsnames]{xcolor}
\usepackage{epstopdf}
\usepackage{epsfig}
\usepackage{comment}
\usepackage{graphicx,color,soul}
\usepackage{multirow}
\usepackage{verbatim}
\usepackage{enumitem}
\usepackage{fixmath}

\usepackage{hyperref}
\usepackage{bold-extra}

\begin{document}


\title{Gauged $\mathbold{L^{}_{\mu}{-}L^{}_{\tau}}$ at a  muon collider}

\author{Guo-yuan Huang}
\email{guoyuan.huang@mpi-hd.mpg.de} 
\affiliation{Max-Planck-Institut f\"ur Kernphysik, Postfach
103980, D-69029 Heidelberg, Germany}

\author{Farinaldo S.\ Queiroz}
\email{farinaldo.queiroz@iip.ufrn.br} 
\affiliation{International Institute of Physics, Universidade Federal do Rio Grande do Norte,
	Campus Universitario, Lagoa Nova, Natal-RN 59078-970, Brazil\\
Departamento de F\'isica, Universidade Federal do Rio Grande do Norte, 59078-970, Natal,
RN, Brasil\\
Millennium Institute for SubAtomic Physics at the High-energy frontIeR, SAPHIR, Chile
}

\author{Werner Rodejohann}
\email{werner.rodejohann@mpi-hd.mpg.de} 
\affiliation{Max-Planck-Institut f\"ur Kernphysik, Postfach
	103980, D-69029 Heidelberg, Germany}

\date{\today}

\begin{abstract}
\noindent
We investigate the sensitivity of the projected TeV muon collider to the gauged $L^{}_{\mu}$-$L^{}_{\tau}$ model. Two processes are considered: $Z'$-mediated two-body scatterings $\mu^+ \mu^-  \to \ell^+ \ell^-$ with $\ell = \mu$ or $\tau$, and scattering with initial state photon emission, $\mu^+ \mu^- \to \gamma Z',~Z' \to \ell \overline{\ell}$, where $\ell$ can be $\mu$, $\tau$ or  $\nu_{\mu/\tau}$. 
We quantitatively study the sensitivities of these two processes by taking into account possible signals and relevant backgrounds in a muon collider experiment with a center-of-mass energy $\sqrt{s} = 3~{\rm TeV}$ and a luminosity $L=1~{\rm ab^{-1}}$.
For  two-body scattering  one can exclude $Z'$ masses $M^{}_{Z'} \lesssim 100~{\rm TeV}$ with $\mathcal{O}(1)$ gauge couplings. 
When $M^{}_{Z'} 
\lesssim 1~{\rm TeV} <\sqrt{s}$, one can exclude $g' \gtrsim 2\times 10^{-2}$. 
The process with photon emission is more powerful than the two-body scattering if $M^{}_{Z'} < \sqrt{s}$. For instance, a sensitivity of $g' \simeq 4 \times 10^{-3}$ can be achieved at $M^{}_{Z'} = 1~{\rm TeV}$. The parameter spaces favored by the $(g-2)^{}_{\mu}$ and $B$ anomalies with $M^{}_{Z'} > 100~{\rm GeV}$ are entirely covered by a muon collider.
\vspace*{3em}
\end{abstract}

\preprint{}

\maketitle
\bigskip

\section{Introduction}
\noindent
Fundamental puzzles in particle physics are the motivation to push the energy frontier with large colliders. 
As a potential future option, the idea to collide muon beams was originally discussed to probe the neutral current in weak interactions~\cite{Tikhonin:2008pw,Budker:1969cd}.
Modern visions of high-energy muon collider experiments are considered to be both energy efficient and cost effective~\cite{Delahaye:2019omf,Shiltsev:2019rfl}. 
Owing to its larger mass, a muon can be accelerated to a considerably higher energy in comparison to an electron, which suffers from the limitation of significant synchrotron radiation. 
Moreover, contrary to hadrons, as an elementary particle the muon is able to deliver all of its energy into a collision. 

Recently, there have been increasing interests in muon colliders at the TeV energy scale~\cite{MCmeeting}, and once being built it would undoubtedly be a powerful window to the Standard Model (SM) and the new physics beyond~\cite{Chiesa:2020awd,Costantini:2020stv,Han:2020pif,Han:2020uak,Bandyopadhyay:2020otm,Gu:2020ldn,Capdevilla:2020qel,Buttazzo:2020eyl,Yin:2020afe}. 

Another advantage of the muon collider is that it can be a perfect probe for muon-philic forces~\cite{Capdevilla:2020qel,Buttazzo:2020eyl,Yin:2020afe}.
There are indeed hints from the anomaly of the muon magnetic moment~\cite{Bennett:2006fi,Roberts:2010cj,Aoyama:2020ynm} that new couplings may exist for muons. 
Indeed, the test of various solutions to the $(g-2)^{}_{\mu}$ anomaly in the muon collider is found to be guaranteed given a reasonable machine setup~\cite{Capdevilla:2020qel}. 
One of the simplest ways to obtain these new interactions is to gauge the accidental $U(1)$ symmetries in SM. 
Among several options, an attractive possibility is the anomaly-free gauged $L^{}_{\mu}$-$L^{}_{\tau}$ model~\cite{He:1990pn,Foot:1990mn,He:1991qd}, whose experimental constraints are relatively loose due to the absence of electron coupling at tree level. Moreover, the flavor structure $L^{}_{\mu}$-$L^{}_{\tau}$ generates attractive parameters in the neutrino sector, namely 
 vanishing $\theta_{13}$, maximal $\theta_{23}$, large $\theta_{12}$ and neutrino masses without a strong hierarchy \cite{Choubey:2004hn,Heeck:2011wj}. 
 An additional motivation of $L^{}_{\mu}$-$L^{}_{\tau}$ came from long-standing anomalies in neutral current $B$ meson decays 
 $B \to K^\ast \mu^+\mu^-$ and the ratio of $B \to K \mu^+\mu^-$ and $B\to K e^+e^-$, see e.g.\  
 \cite{Altmannshofer:2014cfa,Crivellin:2015mga,Altmannshofer:2016jzy}.

\begin{figure*}[t!]
	\begin{center}
			\includegraphics[width=0.7\textwidth]{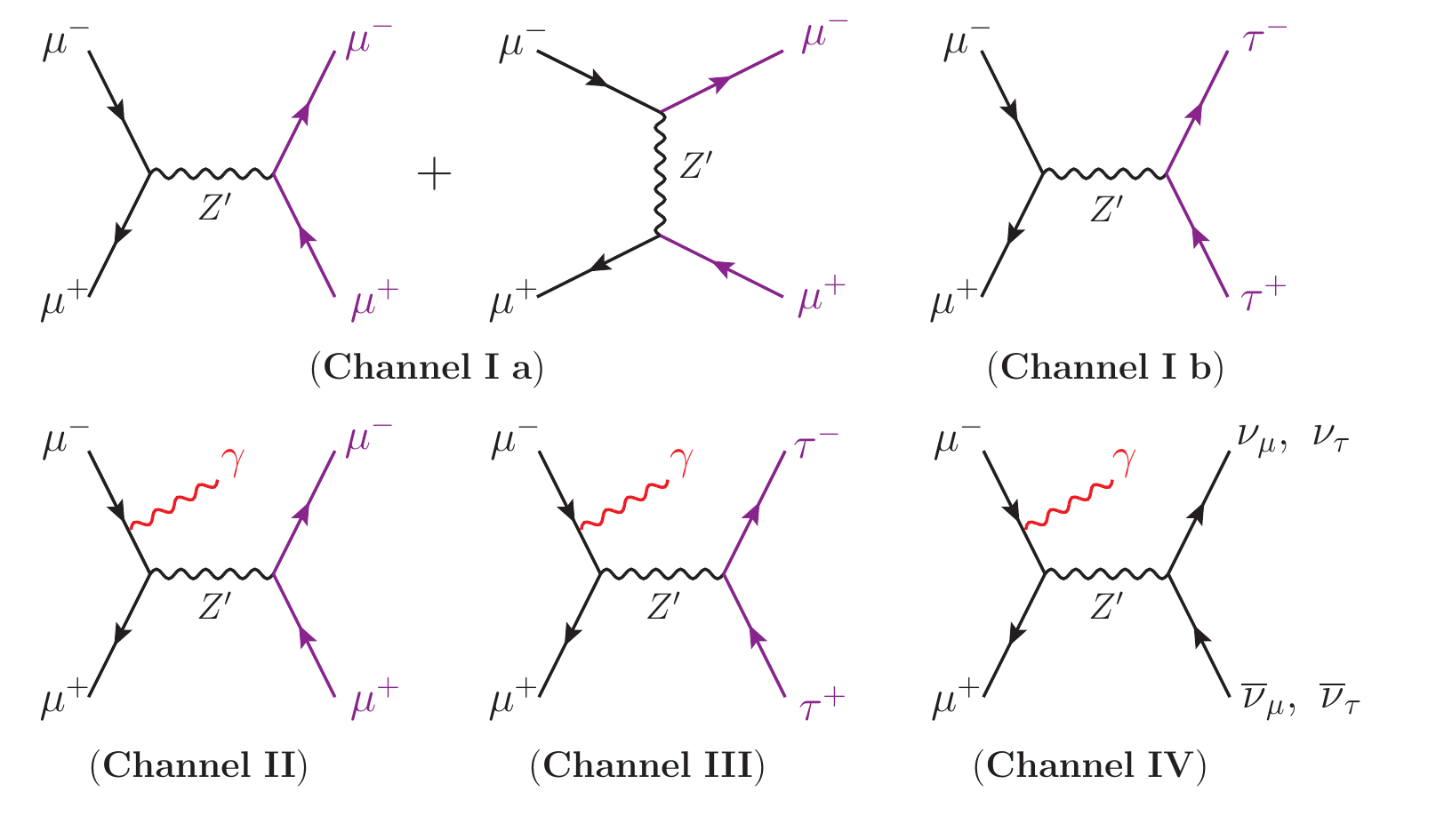} 
	\end{center}
	\vspace{-0.3cm}
	\caption{The leading Feynman diagrams that are sensitive to the gauged $L^{}_{\mu}$-$L^{}_{\tau}$ model in a muon collider. The detectable final states are highlighted in red and purple colors.}
	\label{fig:feynman}
\end{figure*}

We will focus on the phenomenology of the gauged $L^{}_{\mu}$-$L^{}_{\tau}$ model at muon colliders in this work. The gauge interactions with the new boson $Z'$ at tree level are 
\begin{eqnarray} \label{eq:L}
	\mathcal{L} \supset  g^{\prime} \left( \overline{\ell^{}_{\rm L}} Q^{\prime} \gamma^{\mu} \ell^{}_{\rm L} + \overline{E^{}_{\rm R}}Q^{\prime} \gamma^{\mu} E^{}_{\rm R} \right) Z^{\prime}_{\mu}\;,
\end{eqnarray}
where $g'$ is the coupling constant, $\ell \equiv (\nu, E)^{\rm T}$ is the lepton doublet with $\nu$ and $E$ being the neutrino and charged lepton fields, respectively, and $Q^{\prime} = {\rm Diag}(0,1,-1)$ denotes the charge in the flavor basis of $(e,\mu,\tau)$.
The gauge symmetry is required to be broken to obtain a massive $Z'$ particle.
Even though the tree-level coupling for electrons is vanishing,
at loop level  $\gamma$-$Z'$ mixing is generated and will give rise to an effective coupling $g_e$ between electrons and the $Z'$. For large momentum transfer $q^2=M^{2}_{Z'}  \gg M^2_{\tau}$ the result is~\cite{Araki:2017wyg}
\begin{eqnarray} \label{eq:}
 g^{}_{e} \simeq  \frac{2 \alpha g'}{\pi} \frac{m^{2}_{\tau} -m^{2}_{\mu}}{ M^2_{Z'}}   \simeq 1.6 \times 10^{-6} \, g' \left( \frac{100~{\rm GeV}}{M^{}_{Z'}} \right)^2 ,
\end{eqnarray}
which is negligible for our case. This leaves us with the production channels given in Fig.\ \ref{fig:feynman}, to be discussed in detail in the next section. Note that we assume the absence of 
kinetic or mass $Z$-$Z'$ mixing, which makes our limits conservative. 
We will investigate in detail the different channels and their sensitivities, assuming a  muon collider setup with a center-of-mass energy $\sqrt{s} = 3~{\rm TeV}$ and a luminosity $L = 1~{\rm ab^{-1}}$ for illustration. The value $\sqrt{s} = 3~{\rm TeV}$ is one of the options given by the Muon Collider Working Group~\cite{Delahaye:2019omf}, and represents a conservative choice of energy, which according to the muon collider design study~\cite{MCgoal} can be up to 30 TeV. 
The luminosity of $L = 1~{\rm ab^{-1}}$ was given as a minimal value to collect $\mathcal{O}(100)$ events of large-$P^{}_{\rm T}$ $2\to 2$ electroweak scatterings~\cite{Delahaye:2019omf}, which would lead to percent-level measurements of electroweak processes at such high energies. We will demonstrate that the TeV muon collider is highly sensitive to the $L^{}_{\mu}$-$L^{}_{\tau}$ model for  $M^{}_{Z'} $ above $100~{\rm GeV}$, which far exceeds various current limits and  future projections. 
The parameter space for both long-standing anomalies, in the muon's magnetic moment and in $B$ decays, will be fully covered. \\

\begin{figure*}[t!]
	\begin{center}
		\subfigure{
			\includegraphics[width=0.4\textwidth]{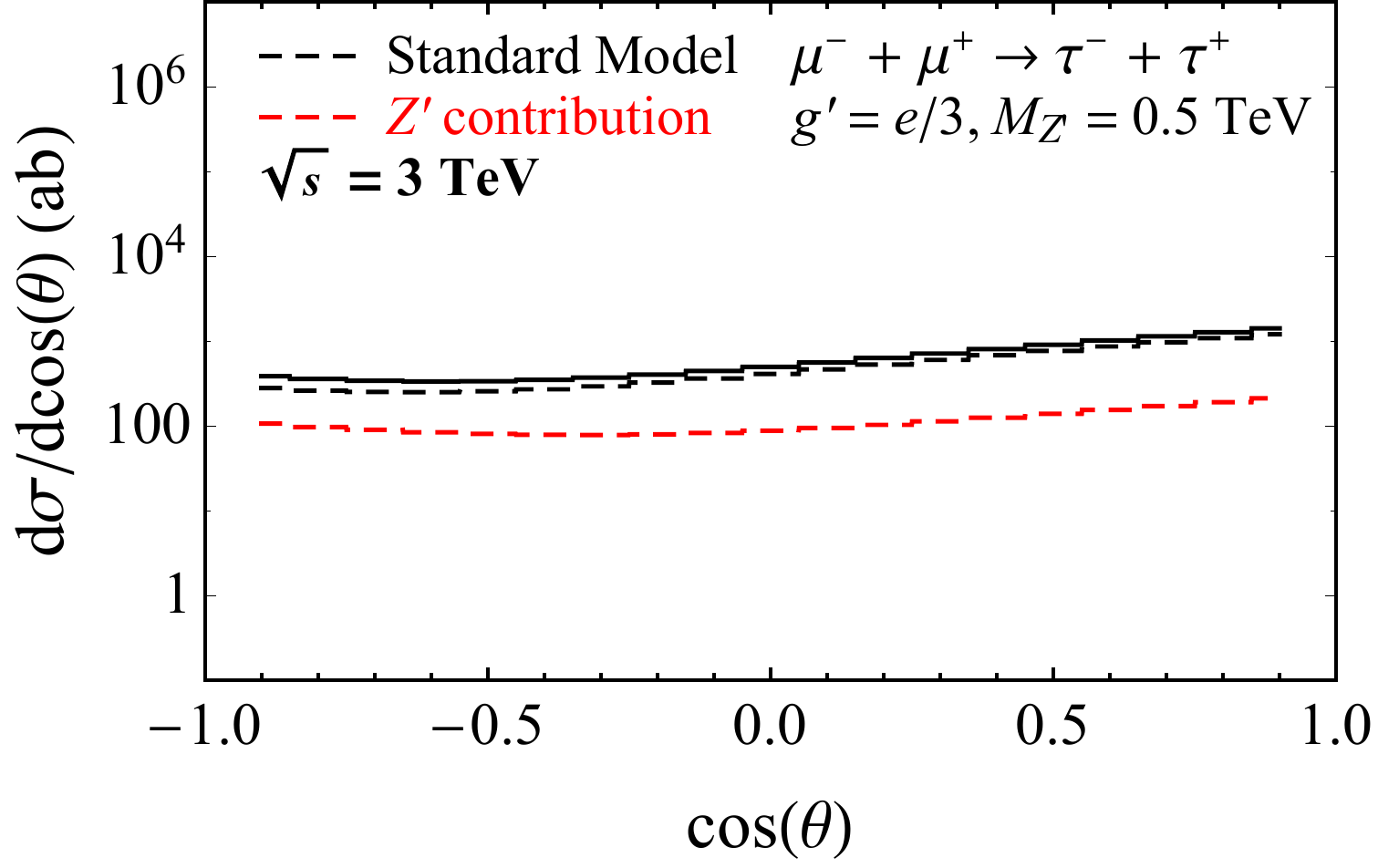} }
		\hspace{1cm}
		\subfigure{
			\includegraphics[width=0.4\textwidth]{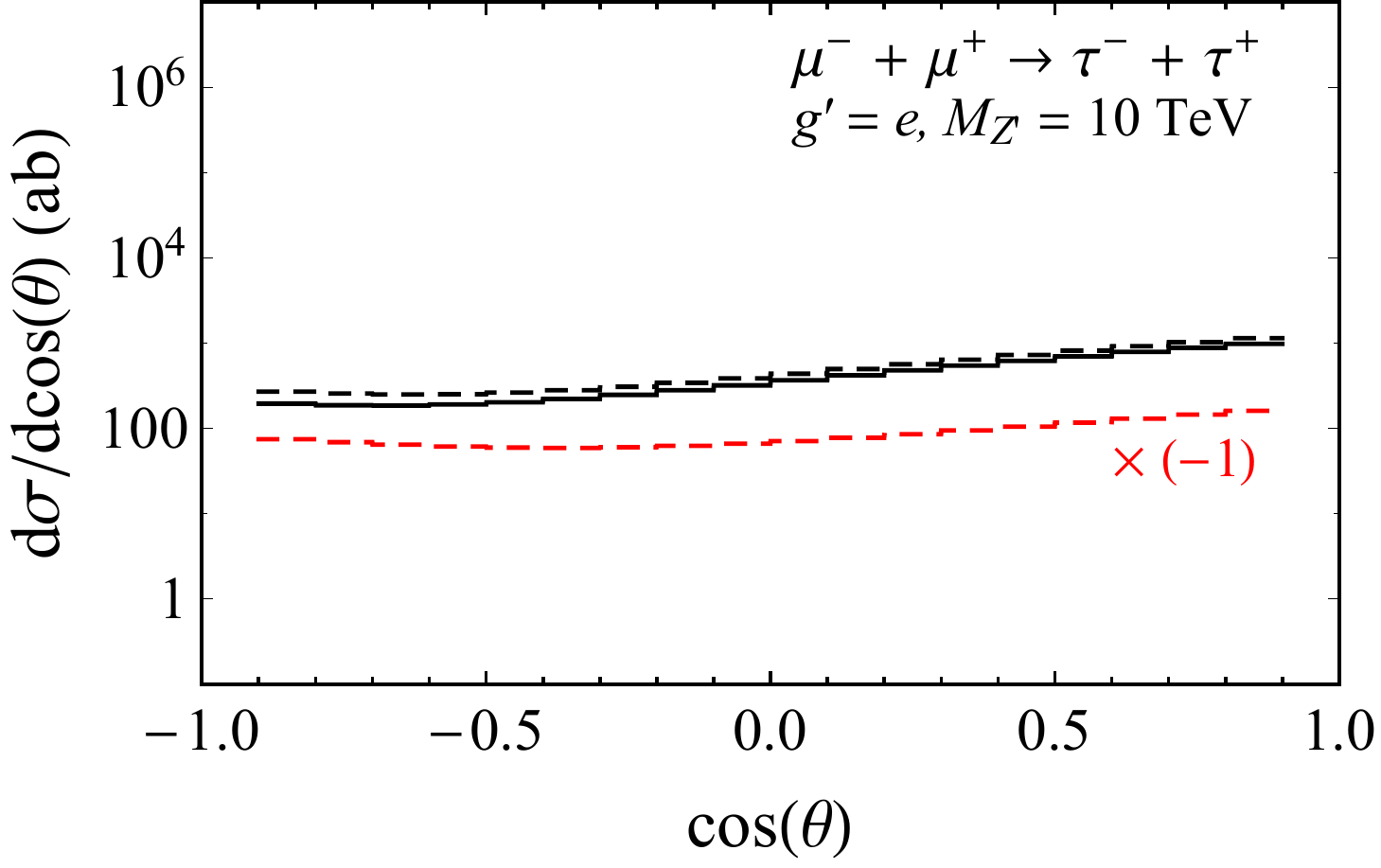} }
		\subfigure{
	\includegraphics[width=0.4\textwidth]{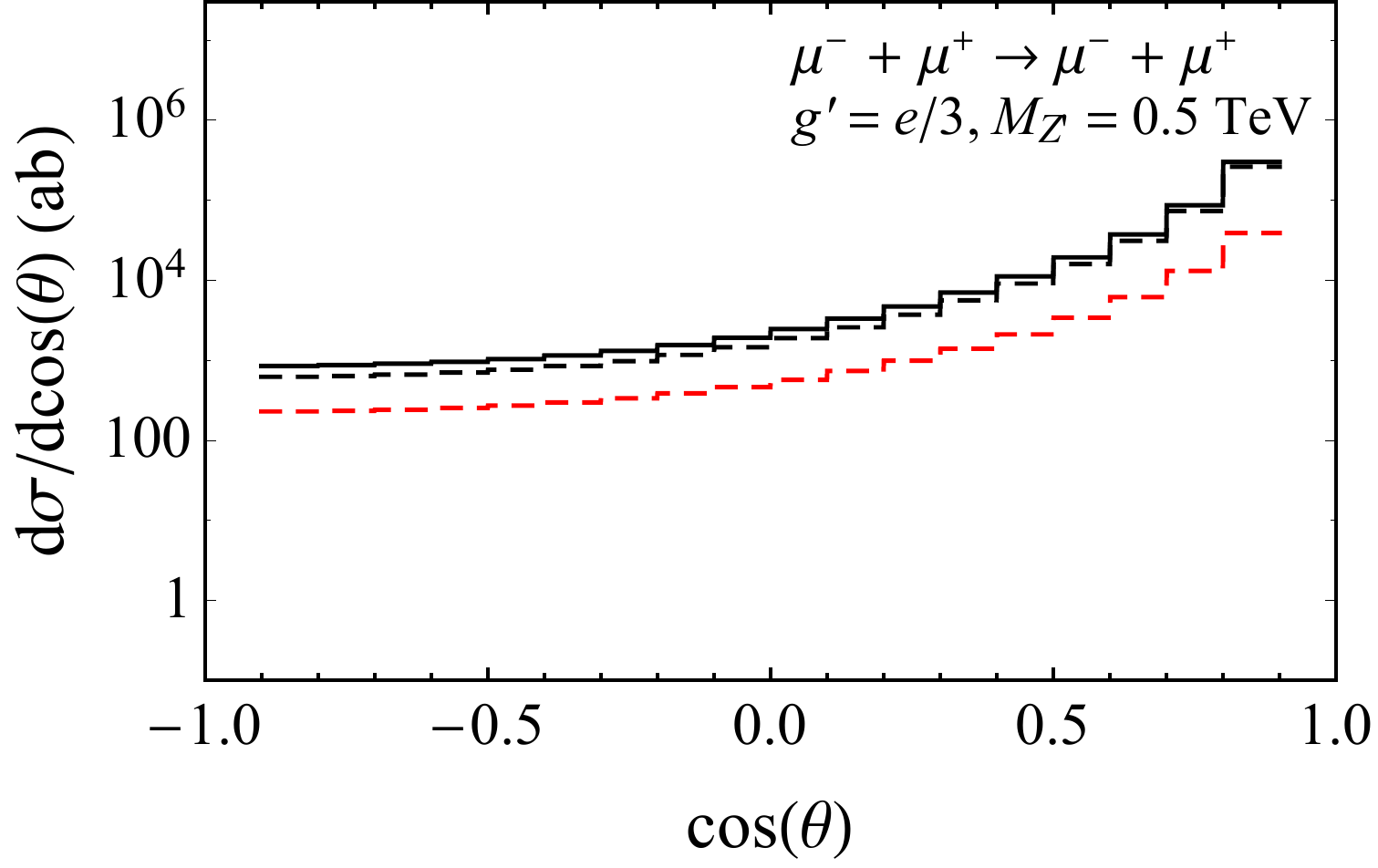} }
		\hspace{1cm}
		\subfigure{
	\includegraphics[width=0.4\textwidth]{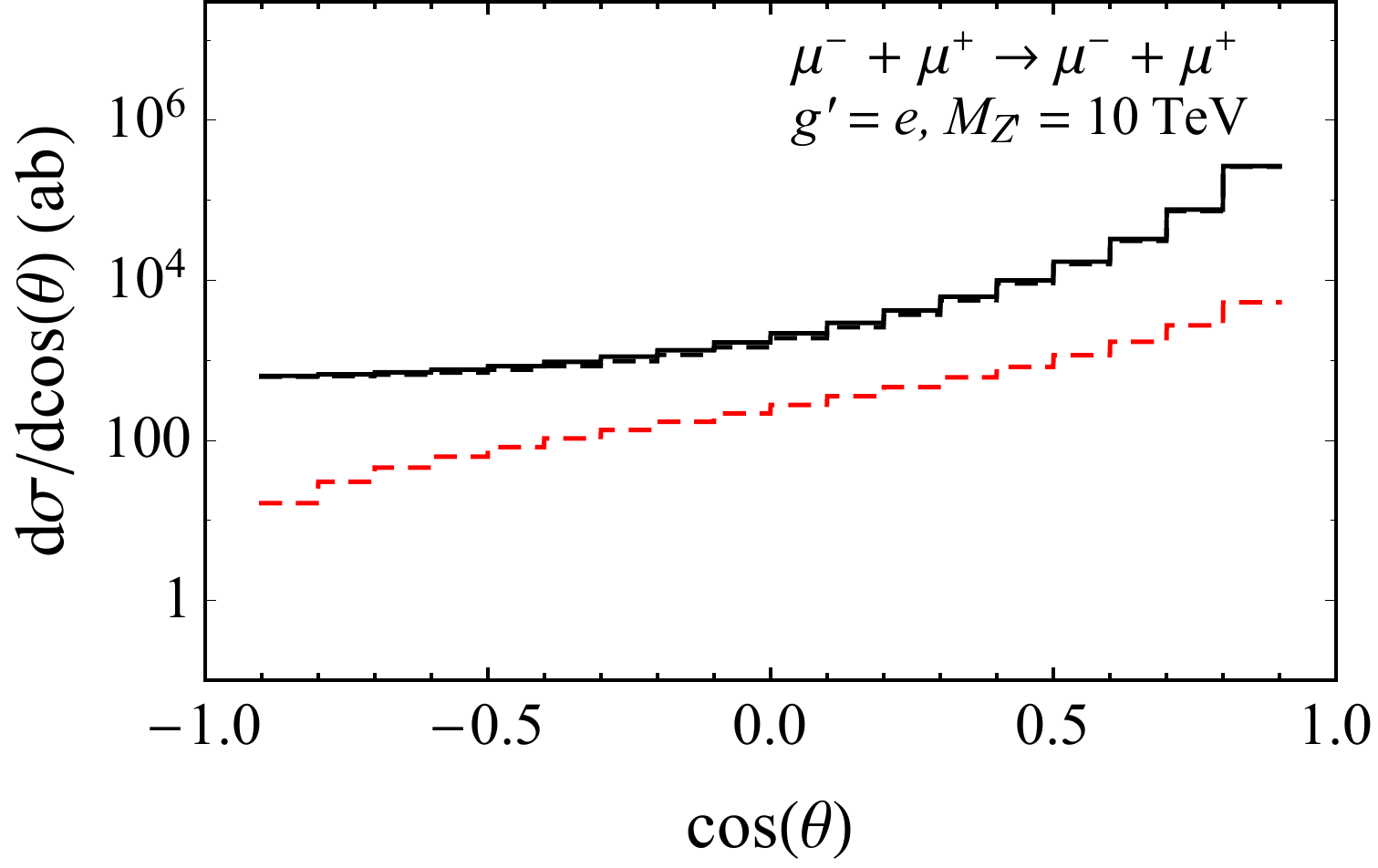} }
	\end{center}
	\vspace{-0.3cm}
	\caption{The differential cross sections as a function of the polar angle $\cos{\theta}$. We have taken $g'=e/3$ and $M^{}_{Z'}=0.5~{\rm TeV}$ for the left two panels and  $g'=e$ and $M^{}_{Z'}=10~{\rm TeV}$ for the right ones. The dashed black histogram stands for the SM backgrounds, and the red ones for the $Z'$ contributions. Their sum is given by the solid black histogram. Note that for the upper-right panel, the $Z'$ contribution is negative, but for clarity we reflect it to the positive axis. 
}
	\label{fig:xsec_mumutautau}
\end{figure*}

 

The rest of this work is organized as follows. In Sec.~\ref{sec:SP}, we list relevant channels which dominate the sensitivities.
The constraining power of the muon collider is examined in Sec.~\ref{sec:CP}, before we conclude in Sec.~\ref{sec:C}.

\section{Scattering processes}\label{sec:SP}
\noindent 
We will focus on two types of scattering processes, as shown in Fig.~\ref{fig:feynman}. 

The first one is the lowest-order process $\mu^+ + \mu^-  \to \ell^+ + \ell^- $ via the exchange of the $Z'$ boson, where the charged lepton $\ell$ can be $\mu$ or $\tau$. 
The signals of $\mu$ and $\tau$ could in principle be separated in the detector by their track multiplicity.
In this process, the massive $Z'$ can manifest itself with two possible effects depending on its mass $M^{}_{Z'}$.
When $M^{}_{Z'}$ is close to $\sqrt{s}$, the resonant scattering significantly enhances  the rate of the signal, enabling also sensitivity to weakly coupled $Z'$.
However, when $M^{}_{Z'}$ is away from $\sqrt{s}$, which is mostly the case in the wide parameter space, the signal would be merely a fluctuation beyond the SM background. 

The second powerful process is at the next order with a photon attached to the initial leg, i.e., $\mu^+ + \mu^- \to \gamma + Z'^{(*)},~ Z'^{(*)} \to \ell + \overline{\ell} $, the so-called radiative return~\cite{Chakrabarty:2014pja,Karliner:2015tga}. 
To have observational effects, here $\ell$ can be either charged leptons or neutrinos. 
An excellent virtue of the process is that the photon will carry away part of the energy of the incoming $\mu$, making the resonant production of $Z'$ viable for $M^{}_{Z'} < \sqrt{s} $.  
In the following we will discuss these two processes in detail.
We use the \texttt{CalcHEP} package~\cite{Pukhov:1999gg,Pukhov:2004ca,Belyaev:2012qa} 
to calculate the scattering amplitudes from the Lagrangian in Eq.~(\ref{eq:L}) and to  integrate the phase space of final states for the numerical results. 

\subsection{Channel I: $\mathbold{\mu^+ + \mu^- \rightarrow l^+ + l^-}$}
For the first scattering process, the back-to-back dimuon final states are produced  through both $s$- and $t$-channel diagrams, whereas only the $s$ channel needs to be taken into account for ditau production. The amplitude is constituted by
\begin{eqnarray} \label{eq:}
\mathcal{M} =  \mathcal{M}^{}_{\gamma} + \mathcal{M}^{}_{Z} + \mathcal{M}^{}_{Z'} \;,
\end{eqnarray}
where the subscript denotes the mediator of the corresponding diagram. After squaring the amplitude, there will be contributions of each single mediator as well as their interference terms.
 
When $g'$ is small compared to the electromagnetic coupling $e = \sqrt{4\pi \alpha}$, the effect of new physics is mostly in the interference between $\gamma$ and $Z'$. Taking for example the scattering $\mu^+ + \mu^- \rightarrow \tau^+ + \tau^-$, the interference term in the total cross section reads
\begin{eqnarray} \label{eq:xsecgzp}
\left(\frac{\mathrm{d} \sigma}{\mathrm{d} \Omega} \right)_{\gamma Z'} = \frac{e^2 g'^2 }{16 \pi^2 s} \cdot \frac{(s - M^{2}_{Z'})(t^2+u^2)}{ s[(s - M^{2}_{Z'})^2 + \Gamma^{2}_{Z'} s^2/ M^2_{Z'}]}
\;,
\end{eqnarray}
which scales as $\sim e^2 g'^2 /s$ for $M^{}_{Z'} \ll \sqrt{s} $.
It can be noticed that depending on the sign of $s-M^2_{Z'}$ in Eq.~(\ref{eq:xsecgzp}), the $Z'$ can lead to either an excess or a deficit of events in comparison to the SM expectation. 
Essentially regardless of the value of $g'$, if $\sqrt{s} \simeq M^{}_{Z'}$ the $Z'$ contribution is governed by the Breit-Wigner resonance 
\begin{eqnarray} \label{eq:}
\left(\frac{\mathrm{d} \sigma}{\mathrm{d} \Omega} \right)_{Z'} = \frac{ g'^4 }{32 \pi^2 s} \cdot \frac{(t^2+u^2)}{ (s - M^{2}_{Z'})^2 + \Gamma^{2}_{Z'} s^2/ M^2_{Z'}}
\;.
\end{eqnarray}
The interference in this case is negligible.
The width is fixed by the decay rate of $Z'$, namely
\begin{eqnarray} \label{eq:width}
 \Gamma^{}_{Z'} = \frac{g'^2 M^{}_{Z'}}{24\pi}\left( N^{}_{\nu} + 2 N^{}_{l}\right)
,
\end{eqnarray}
where $N^{}_{\nu} = N^{}_{l} = 2$ for the gauged $L^{}_{\mu}$-$L^{}_{\tau}$ model are the number of generation of neutrinos and charged leptons coupled to $Z'$ as can be seen in Eq.~(\ref{eq:L}). Note that this expression implies that the scalar sector of the full theory that generates a massive $Z'$ does not lead to observable effects. 

In Fig.~\ref{fig:xsec_mumutautau}, the differential cross sections with respect to the cosine of the polar angle for both $\mu^+ + \mu^- \rightarrow \tau^+ + \tau^-$ and $\mu^+ + \mu^- \rightarrow \mu^+ + \mu^-$ channels are illustrated, where we have taken $g' = e / 3$ and $M^{}_{Z'} = 500~{\rm GeV}$ for the left two panels and $g' = e$ and $M^{}_{Z'} = 10~{\rm TeV}$ for the right ones. The center-of-mass energy as mentioned before is set to $3~{\rm TeV}$ for both panels, so $Z'$ in the given examples is produced off-shell.
In all panels, the dashed black histogram represents the cross section expected in  the SM, while the red one stands for the $Z'$ contribution. The total cross section as a sum of the SM and  $Z'$ contributions is given by the solid black histogram. Note that the angular cut $|\cos{\theta}| < 0.9$ has been taken for illustrative purpose.

For the $s$-channel process $\mu^+ + \mu^- \rightarrow \tau^+ + \tau^-$, a  forward-backward asymmetry
\begin{eqnarray}\label{eq:fb}
A^{}_{\rm FB} = \frac{\sigma(\cos\theta>0) - \sigma(\cos\theta<0)}{ \sigma(\cos\theta>0) + \sigma(\cos\theta<0)} 
\end{eqnarray}
will be induced by the parity-violating nature of the SM $Z$ couplings\footnote{If the initial muon beam was polarized, a parity-violating left-right asymmetry
\begin{eqnarray*}
	A^{}_{\rm LR} = \frac{\sigma(\mu^+ \mu^-_{\rm L}\to \ell^+  \ell^-) - \sigma(\mu^+ \mu^-_{\rm R}\to \ell^+  \ell^-)}{\sigma(\mu^+ \mu^-_{\rm L}\to \ell^+  \ell^-) + \sigma(\mu^+ \mu^-_{\rm R}\to \ell^+  \ell^-)}
\end{eqnarray*}
could be similarly defined, where $\sigma(\mu^+ \mu^-_{\rm L(R)}\to \ell^+  \ell^-)$ refers to the cross section for a left-handed (right-handed) muon scattering on an unpolarized muon.}.
The leading SM contribution to the asymmetry at $\sqrt{s}=3~{\rm TeV}$ is the interference between $Z$ and $\gamma$, whereas the  $Z$-diagram is suppressed by a factor $(1-4\sin^2 \theta_{\rm W})^2 \simeq 0.006$ for $A^{}_{\rm FB}$ 
\cite{PDG}. 
According to the Lagrangian in Eq.~(\ref{eq:L}), for charged leptons the $Z'$-exchange diagram itself is parity conserving, but its interference with the $Z$-exchange diagram will modify the asymmetry. 
We will later perform an analysis using bins of the angular distribution, which implicitly includes the  forward-backward asymmetry.

The $Z'$ contribution to $\mu^+ + \mu^- \rightarrow \tau^+ + \tau^-$, which is approximated as $e^2 g'^2/(4\pi s)$ for $s \gg M^{2}_{Z'}$ and $-e^2 g'^2/(4\pi M^2_{Z'})$ for $s \ll M^{2}_{Z'}$, should be compared to the SM cross section $\sim e^4 /(8\pi s)$. For a given luminosity $L$, if the observed events are in agreement with the SM, one can estimate the $3\sigma$ sensitivity to the 
$Z'$ coupling with a trivial total event comparison: 
\small
\begin{eqnarray} \label{eq:}
	\hspace{-0.3cm}
	g' < 3.4 \times 10^{-2} \left(\frac{\sqrt{s}}{3~{\rm TeV}} \right)^{\frac{1}{2}}  \left(\frac{1~{\rm ab^{-1}}}{L} \right)^{\frac{1}{4}}  {\rm max}\left(1, \frac{M^{}_{Z'}}{\sqrt{s}} \right), 
\end{eqnarray}
\normalsize
where the last factor ${\rm max}(1, {M^{}_{Z'}}/{\sqrt{s}})$ is adopted to make this result applicable to both scenarios, $s \gg M^{2}_{Z'}$ and $s \ll M^{2}_{Z'}$. 
Notice that for the sensitivity in case of  $\sqrt{s} \simeq M^{}_{Z'}$ this expression cannot be applied.

On the other hand, for the scattering process $\mu^+ + \mu^- \rightarrow \mu^+ + \mu^-$, the $t$-channel photon-exchange diagram will significantly enhance the events in the forward direction with a factor $ 1/t^2 = \csc^4(\theta/2)/s^2$.
Also, depending on whether $|t| \gg  M^{2}_{Z'}$ or $|t| \ll  M^{2}_{Z'}$, the 
$Z'$-$\gamma$ interference is proportional to $1/t^2 $ or $1/(t\, M^2_{Z'})$. This analytical observation agrees very well with the bottom two panels in  Fig.~\ref{fig:xsec_mumutautau}. In the left panel with $M^{}_{Z'} = 0.5~{\rm TeV}$, the $Z'$ contribution and the SM background increase with a similar slope as $\cos{\theta}$ approaches one, because both of them are approximately proportional to $1/t^2$. As a consequence, the signal-to-noise ratio is nearly a constant at small polar angles. 
In the right panel with a larger mass $M^{}_{Z'} = 10~{\rm TeV}$, the $Z'$ contribution, proportional to $1/(t\, M^2_{Z'})$ for $M^2_{Z'} > s > t$, increases with a smaller slope compared to the SM one.
In this case, the signal-to-noise ratio will drop significantly when the polar angle is small. 
Note that even though the number of $Z'$ events is large in the forward direction, the final sensitivity  is of course subject to systematic uncertainties in a realistic experimental setup, see below. 

\begin{figure}[t!]
	\begin{center}
		\includegraphics[width=0.4\textwidth]{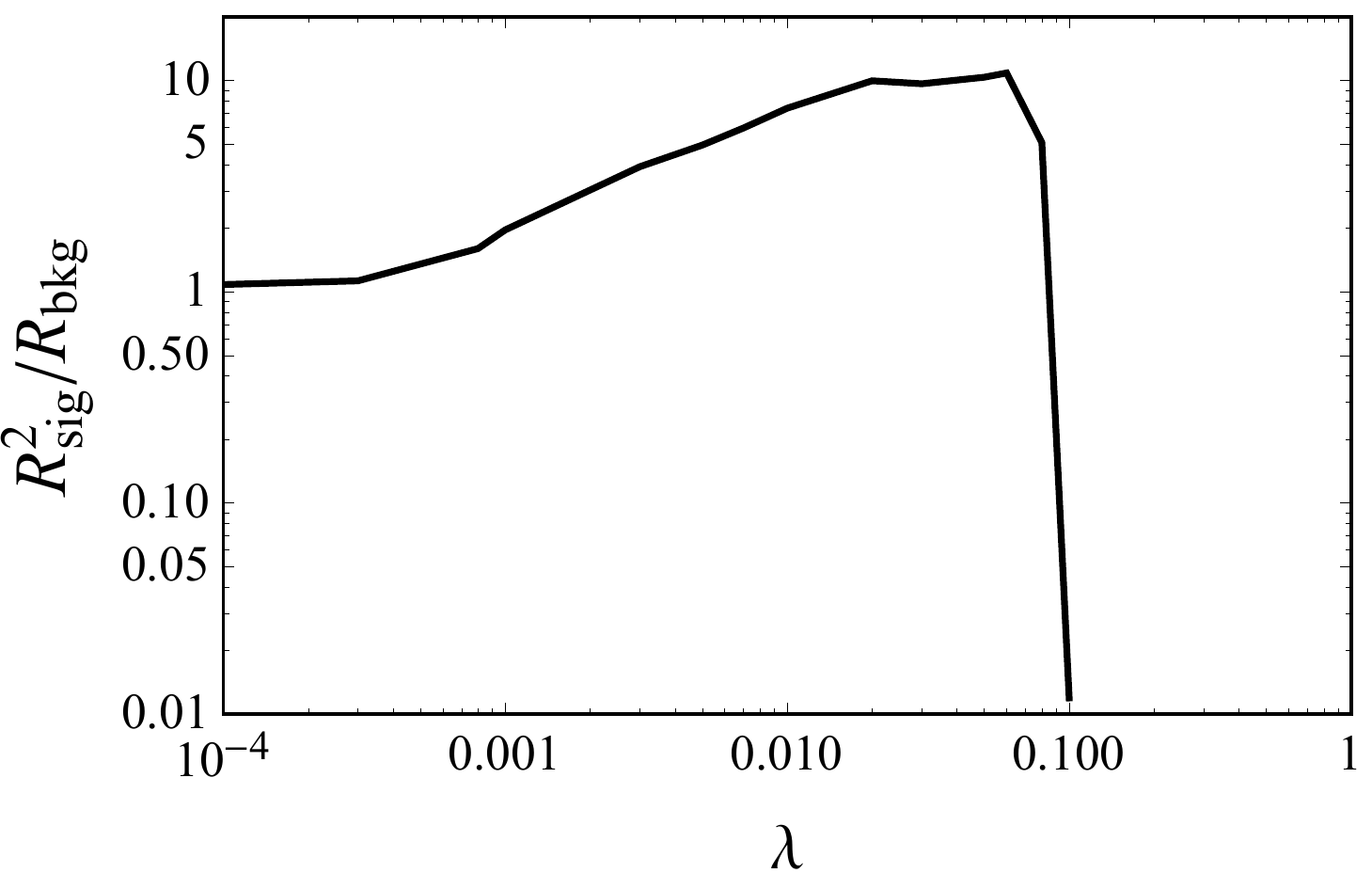} 
	\end{center}
	\vspace{-0.3cm}
	\caption{ $R^{2}_{\rm sig}/R^{}_{\rm bkg}$ as a function of $\lambda$. We take for illustration channel T1a at $m^{}_{\mu\mu} = 1000~{\rm GeV}$. The final $\chi^2$ for the sensitivity is proportional to $R^{2}_{\rm sig}/R^{}_{\rm bkg}$.  }
	\label{fig:cut}
\end{figure}

\subsection{Channel II: $\mathbold{\mu^+ + \mu^- \rightarrow \mu^+ + \mu^- + \gamma}$}
The importance of initial state radiation in searching for the $Z'$ particle is primarily addressed at  electron colliders (see e.g.\  Refs.~\cite{Appelquist:2002mw,Freitas:2004hq,Lees:2014xha,Chakrabarty:2014pja}). 
With a photon attached to one of the initial legs, the center-of-mass energy of the incoming leptons will be effectively reduced, such that the massive intermediate particle can be  produced on-shell, e.g.\ $\mu^+ + \mu^- \to \gamma + Z',~ Z' \to \mu^+ + \mu^- $. This will lead to a resonance peak in the invariant mass spectrum of the  dileptons as well as the equivalent photon energy spectrum.

For the channel $\mu^+ + \mu^- \rightarrow \mu^+ + \mu^- + \gamma$, we are interested in the following two signal typologies:
\begin{itemize}[noitemsep,topsep=1pt,leftmargin=5.5mm]
	\item {\bf T1a}: {\it Dimuon tracks without photon.} 
	Two oppositely charged muon tracks are registered in the detector, while the photon escapes the detection, corresponding to the requirement $|\cos{\theta^{}_{\mu}}| < |\cos{\theta^{}_{\rm det}}|$ and $|\cos{\theta^{}_{\gamma}}| > |\cos{\theta^{}_{\rm det}}|$. Here $\theta^{}_{\rm det} = 10^{\circ} $ is defined as the angular coverage of the detector acceptance following Ref.~\cite{Han:2020uak}.
	The photon produced by the initial state radiation is mainly along the beam direction, for which the cross section is enhanced by a factor of $\ln{\sqrt{s}/m^{}_{\mu}}$; 
	therefore a large part of the signal events registered in the detector should be of this typology.
	The backgrounds are two-fold. One is the same scattering process but with  SM mediators. The other is from $\mu^+ + \mu^- \rightarrow \mu^+ + \mu^- + f^+ + f^-$, where $f=e,\mu,\tau$ escape detection with $|\cos{\theta^{}_{f}}| > |\cos{\theta^{}_{\rm det}}|$. The second background can be dominant in some cases, so it should be included in our calculation. 
	
	The background from $\mu^+ + \mu^- \rightarrow \mu^+ + \mu^- + \gamma$ by $\gamma$- and $Z$-exchange is significantly enhanced by $t$-channel diagrams. In order to reduce this background, we find it is helpful to impose the following kinematic cut
	\begin{eqnarray} \label{eq:tcut}
	\left|t^{}_{\mu^+}\, t^{}_{\mu^-} \right| > 0.01 \, s^2
	\;,
	\end{eqnarray}
	where $t^{}_{\mu^{\pm}} \simeq -\sqrt{s} \, E^{\rm f}_{\mu^{\pm}} (1-\cos{\theta})$ is the squared momentum transfer between the initial and final state leptons, with $E^{\rm f}_{\mu^{\pm}}$ being the final state energy and $
	\theta$ being the scattering polar angle. To justify the choice of the cut, in Fig.\ \ref{fig:cut}, we scan $R^{2}_{\rm sig}/R^{}_{\rm bkg}$ for the cut $|t^{}_{\mu^+}\, t^{}_{\mu^-} | > \lambda \, s^2$ with different $\lambda$ values. Here $R^{}_{\rm sig}$ and $R^{}_{\rm bkg}$ are defined as the ratio of cross section with the cut to that without the cut. The choice of $\lambda = 0.01$ nearly maximizes $\chi^2$ as well as maintains a Gaussian distribution for the background.
	
	\item {\bf T1b}: {\it Dimuon tracks with an accompanying photon}. This typology is similar to the above one, but the photon is required to be registered in the detector with $|\cos{\theta^{}_{\gamma}}| < |\cos{\theta^{}_{\rm det}}|$. Thus only the first background from T1a (i.e., $\mu^+ + \mu^- \rightarrow \mu^+ + \mu^- + \gamma$ with SM mediators) is relevant. Similarly, the kinematic cut in Eq.~(\ref{eq:tcut}) will be imposed to reduce the $t$-channel backgrounds.
\end{itemize}

\begin{figure}[t!]
	\begin{center}
		\includegraphics[width=0.48\textwidth]{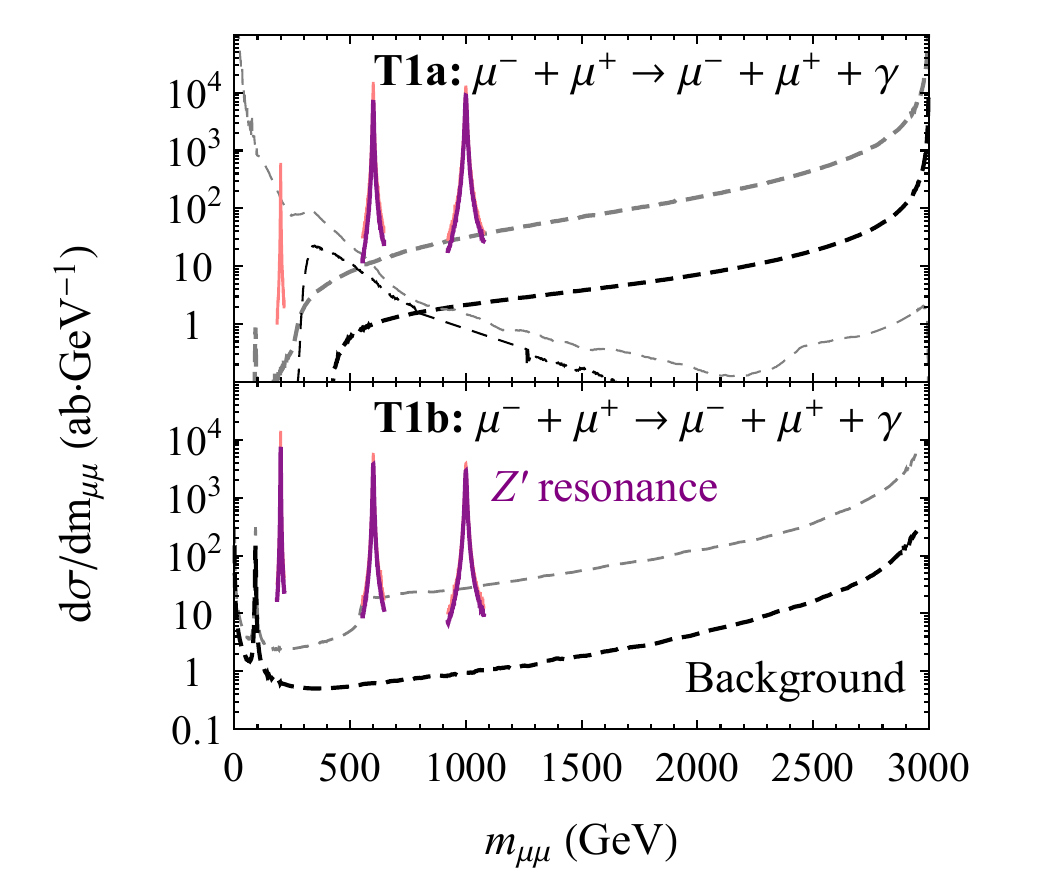} 
	\end{center}
	\vspace{-0.3cm}
	\caption{The invariant mass spectrum of dimuon tracks without photon (upper panel) and with an accompanying photon (lower panel). 
		The purple and black curves stand for the cross sections mediated by $Z'$ and SM particles, respectively, with the kinematic cut $|t^{}_{\mu^+} t^{}_{\mu^-}| > 0.01\, s^2$ to remove the large $t$-channel photon-exchange backgrounds. 
		In comparison, the original cross sections without the kinematic cut are given as the pink and gray curves. 
		From left to right, three peaks correspond to $M^{}_{Z'} = 200,~600~{\rm and }~1000~{\rm GeV}$, respectively. 
		In the upper panel, the thick black curve represents the background of $\mu^+ + \mu^- \to \mu^+ + \mu^- + \gamma$, while the thinner black one is the background $\mu^+ + \mu^- \to \mu^+ + \mu^- + \mu^+ + \mu^-$.}
	\label{fig:typology12}
\end{figure}

In Fig.~\ref{fig:typology12}, we give the invariant mass spectrum of dimuon final states for two typologies {T1a} and {T1b}.
In addition to the kinematic cut in Eq.~(\ref{eq:tcut}), energy cuts  $E^{}_{\mu^{\pm}} >50~{\rm GeV}$ for {T1a}, and $E^{}_{\mu^{\pm},\gamma} >50~{\rm GeV}$ for {T1b} are applied~\cite{Han:2020uak}.
The purple (pink) and black (gray) curves are the $Z'$ signal and background with (without) the  $t$-channel kinematic cut. 
For three resonance peaks, from left to right, $M^{}_{Z'}$ is taken to be $200,~600,~{\rm and }~1000~{\rm GeV}$, respectively.
The $Z'$ coupling has been fixed as  $g'=e$  for concreteness. Taking smaller couplings will shrink the peak width but maintain the peak height. Some observations given below are helpful:
\begin{itemize}[noitemsep,topsep=1pt,leftmargin=5.5mm]
	
	\item It can be clearly noticed that a resonance occurs at the $Z'$ mass of the dimuon spectrum. The peak value exceeds the background by orders of magnitude, which makes the channel very powerful in probing  $Z'$ with feeble couplings.
	
    \item The kinematic cut in Eq.~(\ref{eq:tcut}) is able to reduce the backgrounds dominated by $t$-channel photon exchange by almost two orders of magnitude while decreasing the peak height only by a factor less than two.

    \item For {T1a} (upper panel), the peak height drops dramatically at small $m^{}_{\mu\mu}$. The peak value for $M^{}_{Z'} = 200~{\rm GeV}$ is decreased by more than one order of magnitude compared to other $M^{}_{Z'}$ options. This is in line with our expectation. For smaller $m^{}_{\mu\mu}$, the Lorentz boost effect from the laboratory frame to the dimuon rest frame, which is proportional to $(E^{}_{\mu^{+}} + E^{}_{\mu^{-}})/m^{}_{\mu\mu}$, becomes stronger; therefore two muon tracks with small $m^{}_{\mu\mu}$  will tend to be parallel and follow the opposite direction of the photon, which however is required to escape detection. The additional $t$-channel cut will further reduce the peak, such that it even falls below the range of Fig.~\ref{fig:typology12}. The impact of this reduction on our sensitivity is negligible, as {T1b} will dominate the significance at small $m^{}_{\mu\mu}$ anyways.

\end{itemize}

\subsection{Channel III: $\mathbold{ \mu^+ + \mu^- \rightarrow \tau^+ + \tau^- +  \gamma } $}
The resonant production of a tau pair, $\mu^+ + \mu^- \to \gamma + Z',~Z' \to \tau^+ + \tau^- $, will have nearly the same cross section as the dimuon tracks in the gauged $L^{}_{\mu}$-$L^{}_{\tau}$ model.
However, unlike the muon final state, the energy resolution of taus suffers from the decay into $\nu^{}_{\tau}$. This effect will basically erase the sharp peak of the $Z'$ resonance.

\begin{figure}[t!]
	\begin{center}
		\includegraphics[width=0.48\textwidth]{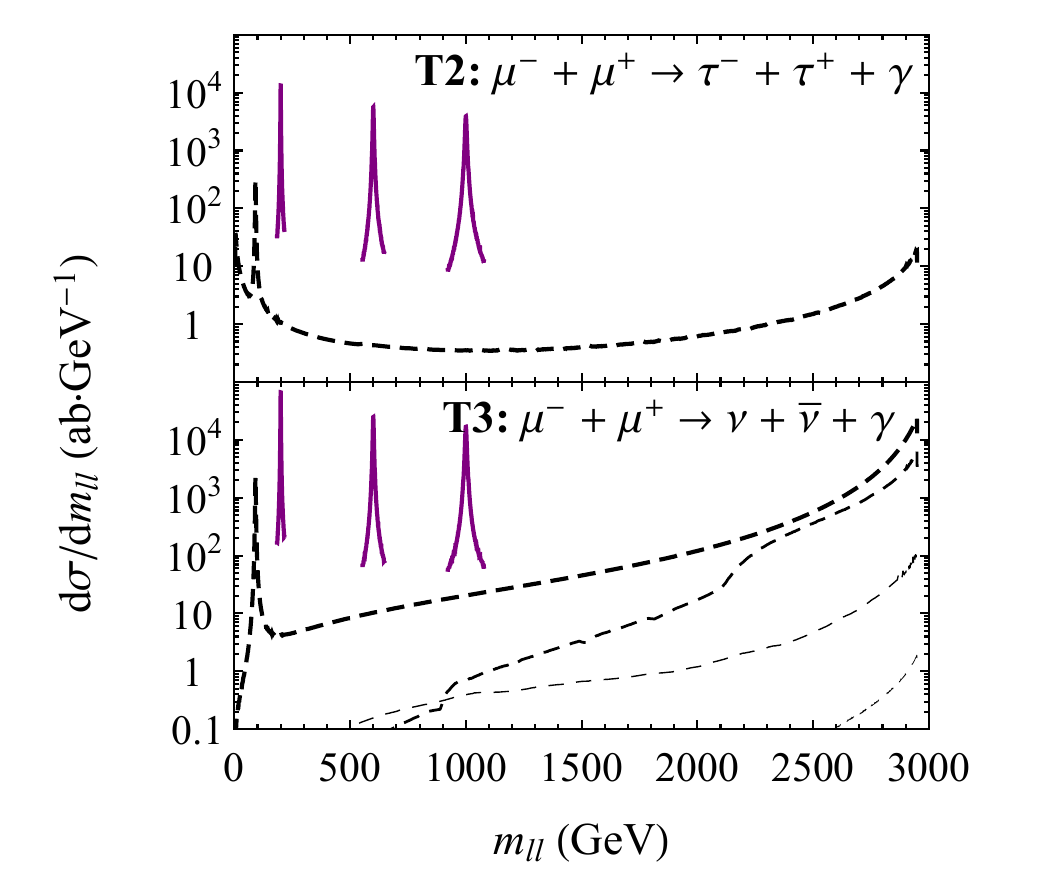} 
	\end{center}
	\vspace{-0.3cm}
	\caption{The invariant mass spectrum of ditau (upper panel) and dineutrino (lower panel) processes. 
		The purple and black curves stand for the cross sections mediated by $Z'$ and SM particles, respectively.  
		From left to right, three peaks correspond to $M^{}_{Z'} = 200,~600~{\rm and }~1000~{\rm GeV}$ respectively. 
		The black curves are the SM backgrounds (see the text for details). }
	\label{fig:typology34}
\end{figure}

In order not to loose  sensitivity, it is necessary to analyze the counter $\gamma$ spectrum instead, and the ditau jets shall be used merely to reject the otherwise large backgrounds. Hence, we  focus on the typology of {\it ditau jets with an accompanying photon} ({\bf T2}) by requiring $|\cos{\theta^{}_{\gamma}}| < |\cos{\theta^{}_{\rm det}}|$ and $|\cos{\theta^{}_{\tau}}| < |\cos{\theta^{}_{\rm det}}|$.
Since there is no $t$-channel background in this scenario, only the energy cuts $E^{}_{\tau} >50~{\rm GeV}$ and $E^{}_{\gamma} >50~{\rm GeV}$ will be imposed for the event generation.

The signal versus background for typology {T2} is shown in the top panel of Fig.~\ref{fig:typology34}. We have taken again $M^{}_{Z'}$ to be $200,~600,~{\rm and }~1000~{\rm GeV}$ for three peaks from left to right. For comparison, the SM background $ \mu^+ + \mu^- \to \gamma + \gamma^* (Z^*),~\gamma^*(Z^*) \to \tau^+ + \tau^- $ is given as the dashed black curve.

\subsection{Channel IV: $\mathbold{\mu^+ + \mu^- \rightarrow \nu + \overline{\nu}+  \gamma }$}
The  {\it monophoton} ({\bf T3}) would be the characteristic signature for the channel $ \mu^+ + \mu^- \rightarrow \nu + \overline{\nu}+  \gamma $, since the  detector is  blind to neutrinos.

In the lower panel of Fig.~\ref{fig:typology34}, we show the differential cross section as a function of the dineutrino invariant mass. The invariant mass is linked to the monophoton energy with a simple relation, namely Eq.~(\ref{eq:ivmgam}) given in the next section.
As before, the purple peaks in the spectrum arise due to the resonant production of $Z'$.
In descending order of magnitude, the backgrounds for monophoton (given as dashed curves) include the following: (i) the contributions of $ \mu^+ + \mu^- \to \nu + \overline{\nu}+  \gamma $ with SM mediators $Z$ and $W$, (ii) the $t$-channel enhanced process $ \mu^+ + \mu^- \to \mu^+ + \mu^- +  \gamma $ where the final $\mu^+ \mu^-$ tracks escape the angular acceptance, (iii) $ \mu^+ + \mu^- \to 3\gamma $ with two photons escaping detection, and (iv) $ \mu^+ + \mu^- \to \ell (q) + \overline{\ell} (\overline{q})+ \gamma $ where $l=e,\tau$ and $q$ are charged leptons and quarks,  respectively. The first contribution $ \mu^+ + \mu^- \to \nu + \overline{\nu}+  \gamma $  dominates the backgrounds, though $ \mu^+ + \mu^- \to \mu^+ + \mu^- +  \gamma $ can make a considerable contribution for large $m^{}_{\mu\mu}$. The other contributions are essentially negligible.

\section{Constraining power} \label{sec:CP}
\noindent 
The actual sensitivity of the process $\mu^+ + \mu^- \to \ell^+ + \ell^-$ (Channel I) also depends on the detection efficiency for the final dilepton pairs. 
Muons will be identified as a prompt isolated track, while taus have  multiple possible decay signals: muons with missing energy for the leptonic decay, and 1-prong and 3-prong pion tracks for the hadronic decays.
The selection efficiency at the $Z$ pole of past electron colliders is large~\cite{ALEPH:2005ab}, e.g., above $95\%$ for muons and $70\%$ for taus.
The design target of future electron colliders is a nearly $100\%$ track identification efficiency~\cite{CEPCStudyGroup:2018ghi,Charles:2018vfv}, which allows an excellent efficiency for muons. 
However, the tau efficiency is limited by the hadronic backgrounds.
An estimation of the Higgs to ditau measurement at future electron colliders gives an overall efficiency of $80\%$ for the $qqH$ channel~\cite{Yu:2020bxh}.
In the following, we will assume the efficiency for dimuon detection to be $100\%$ and that for ditau detection to be conservatively $70\%$.
The results for a lower or higher efficiency could in principle be obtained by rescaling the chi-square. We will adopt the following chi-square form:
\begin{eqnarray} \label{eq:}
	\chi^2_{\rm I} = \sum^{}_{i}\frac{(N^{}_{i}-\widetilde{N}^{}_{i})^2}{N^{}_{i} + \epsilon^2 \cdot N^{2}_{i}}
	\;,
\end{eqnarray}
where $\epsilon$ denotes the systematic uncertainty, which will be taken as  $0.1\%$~\cite{Han:2020uak}, $N^{}_{i}$ is the expected event number of signal plus background, and $\widetilde{N}^{}_{i}$ stands for the experimental observations. Here $i$ sums over bins of polar angles, and we set the bin size of $\cos{\theta}$ as $0.1$. Hence, the information of the spectrum shape (e.g., the forward-backward asymmetry from Eq.\ (\ref{eq:fb})) is contained in $\chi^2_{\rm I}$.

On the other hand, the sensitivity of the process $\mu^+ + \mu^- \to  \ell + \overline{\ell} + \gamma$ relies on the observation of a bump above the background in the invariant mass spectrum of the dileptons.
The width of the resonance is very narrow for weak $Z'$ couplings, such that the smearing effect of finite energy resolution will become crucial.
For Channel II $\mu^+ + \mu^- \rightarrow \mu^+ + \mu^- + \gamma$, there are two approaches to the determination of peak in the invariant mass spectrum: (i) the direct measurement of dimuon tracks, and (ii) the detection of the photon partner.
They are equivalent except for different resolution powers on the invariant mass.
For the general process $\mu^+ + \mu^- \to  \ell + \overline{\ell} + \gamma$, the photon energy is connected to the invariant mass $m^{}_{\ell\overline{\ell}}$ via
\begin{eqnarray} \label{eq:ivmgam}
E^{}_{\gamma} = \frac{s-m^2_{\ell\overline{\ell}}}{2\sqrt{s}}
\;,
\end{eqnarray}
so we have the relation $|\Delta E^{}_{\gamma}| = |\Delta m^{}_{\mu^+\mu^-}| \cdot (m^{}_{\mu^+\mu^-}/\sqrt{s})$, where $|\Delta E^{}_{\gamma}|$ and $|\Delta m^{}_{\mu^+\mu^-}|$ are the resolutions of the detector.
The direct dimuon track measurement is subject to the inaccurate momentum resolution,  and we will assume the form $\Delta m^{}_{\mu^+\mu^-} \simeq 5\times 10^{-5}~{\rm GeV^{-1}}\cdot s$~\cite{Freitas:2004hq}.
For the detection of photons, the electromagnetic calorimeter can yield excellent resolution of energy. For illustration, we will adopt the energy resolution as in the current CMS detector with ${\rm PbWO}^{}_{4}$ crystals~\cite{Chatrchyan:2008aa}:
\begin{eqnarray} \label{eq:}
\frac{\Delta E^{}_{\gamma}}{E^{}_{\gamma}} = \sqrt{\left(\frac{2.8 \%}{\sqrt{E^{}_{\gamma}}}\right)^2+\left( \frac{0.12}{E^{}_{\gamma}}\right)^2 + \left(0.3\%\right)^2} 
\;.
\end{eqnarray}
We can convert this resolution into that of $m^{}_{\mu^-\mu^+}$ according to Eq.~(\ref{eq:ivmgam}).
A comparison between the resolution of invariant mass spectrum based on two different approaches is made, and we find that the photon detection can always give a much better $m^{}_{\mu^+\mu^-}$ resolution. Thus, for typologies {T1b}, {T2} and {T3}, we will adopt the photon energy resolution. For typology {T1a}, however, the momentum measurement of tracks is the only option.

\begin{figure*}[t!]
	\begin{center}
		\hspace{-1cm}
		\subfigure{
			\includegraphics[width=0.45841\textwidth]{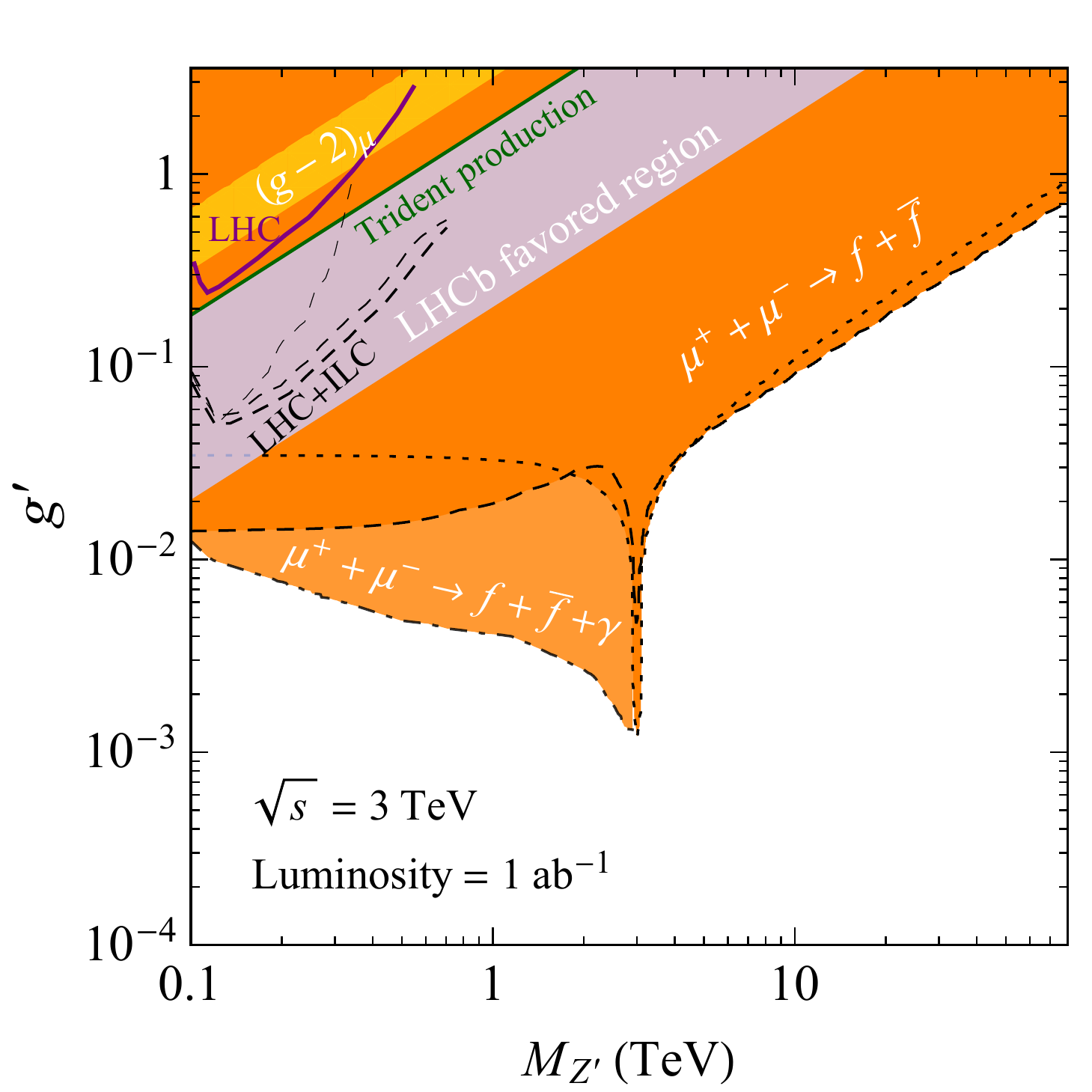} }
		\hspace{1cm}
		\subfigure{
			\includegraphics[width=0.45841\textwidth]{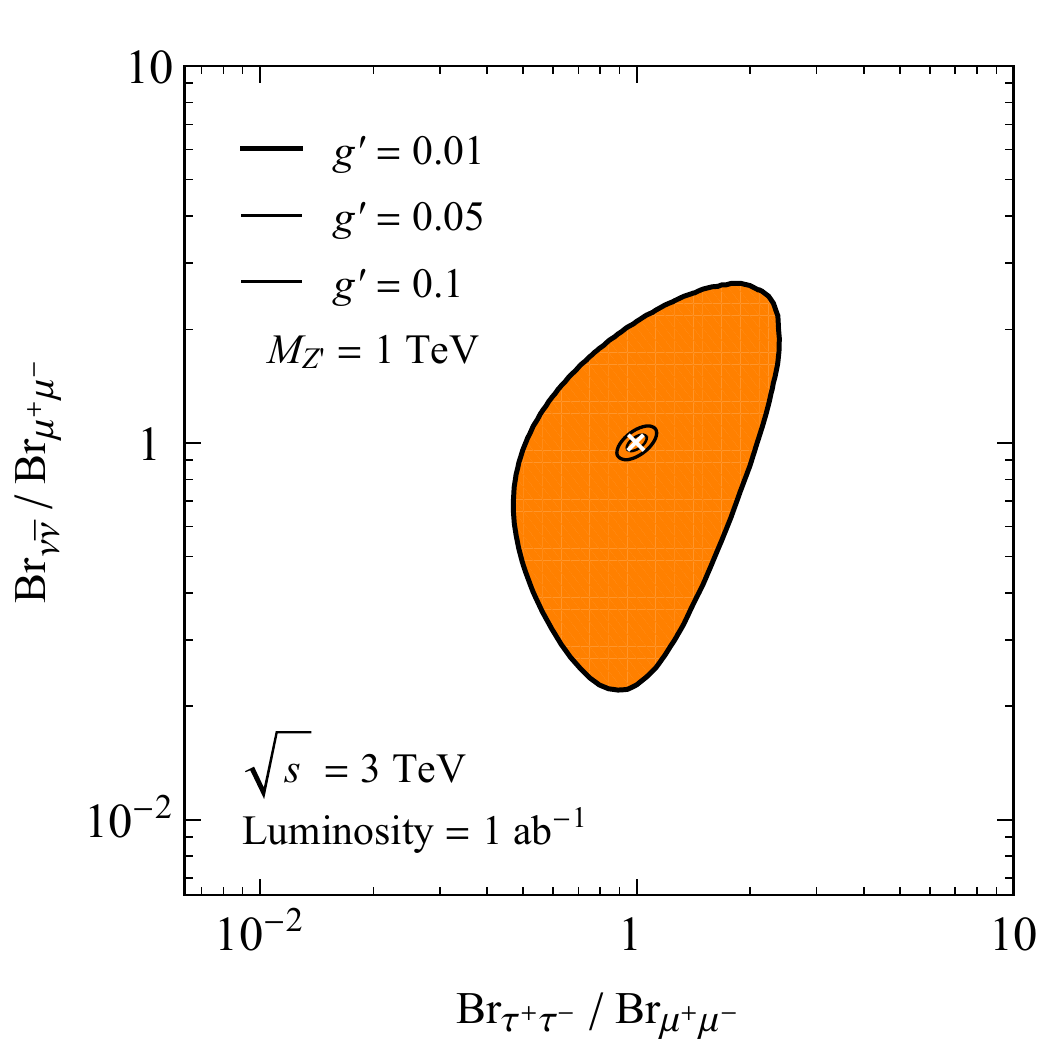} }
	\end{center}
	\vspace{-0.3cm}
	\caption{The sensitivity of the muon collider with a center-of-mass energy $\sqrt{s}=3~{\rm TeV}$ and luminosity $1~{\rm ab^{-1}}$, with SM background observed only (left panel) and with a $Z'$ signal (right panel). In the left panel, the orange region is derived by assuming that only the SM background is observed, i.e., no excess of events. The darker orange region is obtained from Channel I,  $\mu^+ + \mu^- \rightarrow \ell^+ + \ell^-$, and the lighter orange region is from Channels II, $\mu^+ + \mu^- \rightarrow \mu^+ + \mu^- + \gamma$, III $\mu^+ + \mu^- \rightarrow \tau^+ + \tau^- + \gamma$, and IV $\mu^+ + \mu^- \rightarrow \nu + \overline{\nu} + \gamma$. The parameter space favored by the $(g-2)^{}_{\mu}$ anomaly by $2\sigma$ is shown in the yellow band, and the region supported by the $B$ anomalies is reproduced as the light blue band~\cite{Altmannshofer:2016jzy}.
	The neutrino trident production limit is given as the green curve~\cite{Altmannshofer:2014pba}, and the LHC bound recast from $3\ell$ data is shown as the purple curve~\cite{Drees:2018hhs}. The projection of LHC and ILC with different channels is given in three dashed black curves~\cite{delAguila:2014soa}.
	In the right panel, we choose $M^{}_{Z'} = 1~{\rm TeV}$ to generate the experimental events. From outer to inner contours, $g'$ is taken to be $0.01$, $0.05$ and $0.1$, respectively. The white cross in the center of the contours is the prediction of the gauged $L^{}_{\mu}$-$L^{}_{\tau}$ model.}
	\label{fig:contour}
\end{figure*}

We will first examine the sensitivity by deriving constraints based on the following considerations. 
First, we assume only the background is observed, such that an upper bound can be placed on the $Z'$ coupling $g'$ for each given $M^{}_{Z'}$.
Second, for $\mu^+ + \mu^- \to  \ell + \overline{\ell} + \gamma$, the half width of the region of interest in the invariant mass spectrum is chosen to be the maximal one among  the peak width and $m^{}_{\ell\overline{\ell}}$ resolution. Third, the significance of Channels II ($\mu^+ + \mu^- \rightarrow \mu^+ + \mu^- + \gamma$), III ($\mu^+ + \mu^- \rightarrow \tau^+ + \tau^- + \gamma$), and IV ($\mu^+ + \mu^- \rightarrow \nu + \overline{\nu} + \gamma$) can be obtained by combining all typologies
\begin{eqnarray} \label{eq:}
\chi^2_{\rm II,III,IV} = \frac{(N^{}_{\rm ROI}-\widetilde{N}^{}_{\rm ROI})^2}{ N^{}_{\rm ROI} +\epsilon^2 \cdot N^{2}_{\rm ROI}}
\;,
\end{eqnarray}
where $\widetilde{N}^{}_{\rm ROI}$ is the number of observed events within the region of interest and $N^{}_{\rm ROI}$ is the theoretical expectation given a $Z'$ model.


The prospective limits obtained in this way  need to be compared with current constraints. 
Existing bounds on the gauged $L^{}_{\mu}$-$L^{}_{\tau}$ model span a wide range of the $Z'$ mass $M^{}_{Z'}$.
For $M^{}_{Z'}$ around  GeV, the BaBar experiment \cite{TheBABAR:2016rlg} with the channel $e^{+} + e^{-} \to \mu^{+} + \mu^{-} + Z',~Z' \to \mu^{+} + \mu^{-} $ can place the most stringent limits, e.g., $g^{}_{Z'} \lesssim 10^{-3}$ at $M^{}_{Z'} \simeq 1~{\rm GeV}$.
The LHC searches dominate the $Z'$ bounds above a few GeV~\cite{Ma:2001md,Heeck:2011wj,Aad:2014wra,Elahi:2015vzh,Altmannshofer:2016jzy,Sirunyan:2018nnz,Drees:2018hhs,delAguila:2014soa,Nomura:2020vnk}.
In particular, a dedicated search for $Z'$ by the CMS collaboration with $Z \to 4\mu$ events can set the most competitive constraints in the mass range $5~{\rm GeV} \lesssim M^{}_{Z'} \lesssim 60~{\rm GeV}$~\cite{Sirunyan:2018nnz}. 
In a wider range, $Z'$ bounds have also been recast from other measurements at LHC~\cite{Drees:2018hhs}.
Further above, the constraints from trident production in neutrino scattering experiments, $\nu^{}_{\mu} + N \to \nu^{}_{\mu} + \mu^+ + \mu^- + N$, are the most stringent ones with the limit $g' \lesssim M^{}_{Z'}/(540~{\rm GeV})$ for decoupled $Z'$ masses~\cite{Altmannshofer:2014pba}. However, these existing bounds become rather weak above a few hundreds of GeV, mainly owing to the exclusive muonic and taunic 
couplings. 

The projected bounds obtained in this paper are presented in the left panel of Fig.~\ref{fig:contour}. The limits using Channel I, $\mu^+ + \mu^- \rightarrow \ell^+ + \ell^-$, are given as the darker orange region, where the dashed curve stands for the case of $\ell=\mu$ and the dotted curve for $\ell=\tau$. The lighter orange region shows the constraining power of processes with initial state radiation, namely Channels II, III and IV.
One can notice that these channels\footnote{The resonant process $\mu^+ + \mu^- \to \gamma + Z'$ and the new physics contribution of two-body scattering $\mu^+ + \mu^- \to \ell^+ + \ell^-$ are actually at the same order.} have a much better sensitivity than the two-body scattering ones for $M^{}_{Z'} < \sqrt{s}$, even though the signal rates are both proportional to $e^2 g'^2 $. This should be ascribed to the fact that the initial state radiation is enhanced by a factor of $\ln{(\sqrt{s}/p^{\rm T,cut}_{\gamma})}$ with $p^{\rm T,cut}_{\gamma}$ being the transverse momentum of the  photon within the detector acceptance.

For comparison, the parameter spaces explaining the $(g-2)^{}_{\mu}$ and $B$ anomalies~(taken from \cite{Altmannshofer:2016jzy}) are given as the yellow and blue bands, respectively. Various existing bounds and projections shown in the plot include: the limit of neutrino trident production (green curve)~\cite{Altmannshofer:2014pba}, a  recast LHC bound (purple curve)~\cite{Drees:2018hhs}, and the exclusion projection of LHC and ILC with various channels (dashed black curves)~\cite{delAguila:2014soa}. For $M^{}_{Z'} > 100~{\rm GeV}$, the parameter space favored by the $B$ anomalies is entirely covered by the projection of muon collider.

On the contrary, if the $L^{}_{\mu}$-$L^{}_{\tau}$ model is true, one would measure its parameters instead of constraining them. In particular, if a $Z'$ signal manifests itself in the future muon collider, one may wonder if it is indeed a gauged $L^{}_{\mu}$-$L^{}_{\tau}$ model or some other $Z'$ model. This can be achieved by comparing the branching ratios of the different final state dileptons for $\mu^+ + \mu^- \to \gamma+ Z',~Z' \to \ell + \overline{\ell} $. For instance, for the  $L^{}_{\mu}$-$L^{}_{\tau}$ model, we should have ${\rm Br}^{}_{\mu^+ \mu^-} = {\rm Br}^{}_{\tau^+ \tau^-} = {\rm Br}^{}_{\nu \overline{\nu}}$, where ${\rm Br}^{}_{f \overline{f}}$ denotes the branching ratios for $Z'$ decays to $f \overline{f}$. In the right panel of Fig.~\ref{fig:contour}, we show the $2\sigma$ allowed region for $ {\rm Br}^{}_{f \overline{f}} /{\rm Br}^{}_{\mu^+ \mu^-}$ assuming $L^{}_{\mu}$-$L^{}_{\tau}$ is true, with $\sqrt{s}=3~{\rm TeV}$, $L=1~{\rm ab^{-1}}$ and $M^{}_{Z'}=1~{\rm TeV}$. The gauge coupling has been taken $g' = 0.01,\, 0.05,\, 0.1$. The white cross marks the expectation for the $L^{}_{\mu}$-$L^{}_{\tau}$ model. For $g' = 0.1$, $ {\rm Br}^{}_{\tau^+\tau^-} /{\rm Br}^{}_{\mu^+ \mu^-}$ can be confined in the range $(0.095,1.057)$. 
More data could be collected with a larger coupling, such that the allowed region will shrink towards the true case of $L^{}_{\mu}$-$L^{}_{\tau}$ for larger $g'$. 

\section{Conclusion}\label{sec:C}
\noindent
We have studied the sensitivity of a possible muon collider to the well-motivated gauged $L^{}_{\mu}$-$L^{}_{\mu}$ model, assuming a benchmark collider setup of $\sqrt{s} = 3~{\rm TeV}$ and $L = 1~{\rm ab^{-1}}$. Electrons can be millicharged under the $L^{}_{\mu}$-$L^{}_{\tau}$ gauge group through $Z'$-$\gamma$ mixing, which however is negligible for our high energy scenario. We identify two powerful processes: one is the trivial two-body scattering $\mu^+ + \mu^- \rightarrow \ell^+ + \ell^-$ with $\ell = \mu$ or $\tau$, and the other is $\mu^+ + \mu^- \rightarrow \ell + \overline{\ell} + \gamma$ where $\ell$ can be $\mu$ or $\tau$ or $\nu_{\mu/\tau}$.
For the first process, the effect of $Z'$ is a distortion in the final state lepton  angular spectrum, and the $Z'$ coupling needs to be sizable to exceed the SM backgrounds.
In the second process, $Z'$ will manifest itself as a resonance bump above the SM backgrounds, which turns out to be sensitive to $Z'$ with very weak couplings. 

The parameter space of $Z'$ masses above 100 GeV explaining the anomalous magnetic moment of the muon, as well as the $B$ meson anomalies, is fully covered by the muon collider setup, adding further motivation to the facility.


\begin{acknowledgments}
\noindent
GYH was supported by the Alexander von Humboldt Foundation. FSQ is supported by the S\~{a}o Paulo Research Foundation (FAPESP) through grant 2015/15897-1 and ICTP-SAIFR FAPESP grant 2016/01343-7. FSQ acknowledges support from CNPq grants 303817/2018-6 and 421952/2018-0 and the Serrapilheira Institute (grant number Serra-1912-31613).
\end{acknowledgments}


\bibliographystyle{utcaps_mod}
\bibliography{references}

\clearpage

\end{document}